\documentclass[sigconf]{acmart}

\usepackage{amsmath,amsfonts}
\usepackage{algorithmic}
\usepackage{graphicx}
\usepackage{textcomp}
\usepackage{xcolor}
\usepackage[linesnumbered,ruled,vlined]{algorithm2e}
\usepackage{bm}

\usepackage{booktabs} 
\usepackage{comment}
\usepackage{multirow}
\usepackage{url}
\usepackage{amsmath}
\usepackage{bbold}
\usepackage{balance}

\DeclareMathOperator*{\argmin}{arg\,min}

\newcommand{\squishlist}{
 \begin{list}{$\bullet$}
  { \setlength{\itemsep}{0pt}
     \setlength{\parsep}{3pt}
     \setlength{\topsep}{3pt}
     \setlength{\partopsep}{0pt}
     \setlength{\leftmargin}{1.5em}
     \setlength{\labelwidth}{1em}
     \setlength{\labelsep}{0.5em} } }

\newcommand{\squishlisttwo}{
 \begin{list}{$\bullet$}
  { \setlength{\itemsep}{0pt}
     \setlength{\parsep}{0pt}
    \setlength{\topsep}{0pt}
    \setlength{\partopsep}{0pt}
    \setlength{\leftmargin}{2em}
    \setlength{\labelwidth}{1.5em}
    \setlength{\labelsep}{0.5em} } }

\newcommand{\squishend}{
  \end{list}  }
\newcommand{\etal}{\textit{et al}. }
\AtBeginDocument{%
  \providecommand\BibTeX{{%
    \normalfont B\kern-0.5em{\scshape i\kern-0.25em b}\kern-0.8em\TeX}}}

\begin{document}
\title{A Model of Two Tales: Dual Transfer Learning Framework for Improved Long-tail Item Recommendation}


\author{Yin Zhang*, Derek Zhiyuan Cheng, Tiansheng Yao, Xinyang Yi, Lichan Hong, Ed H. Chi}
\thanks{*Work done while interning at Google.}
\email{zhan13679@tamu.edu, {zcheng, tyao, xinyang, lichan, edchi}@google.com}
\affiliation{Google, Inc  \country{USA} \\ Texas A\&M University \country{USA}}

\renewcommand{\shortauthors}{Yin Zhang, et al.}

\begin{abstract}
Highly skewed long-tail item distribution is very common in recommendation systems. It significantly hurts model performance on tail items. To improve tail-item recommendation, we conduct research to transfer knowledge from head items to tail items, leveraging the rich user feedback in head items and the semantic connections between head and tail items. Specifically, we propose a novel dual transfer learning framework that jointly learns the knowledge transfer from both model-level and item-level: 1. The model-level knowledge transfer builds a generic meta-mapping of model parameters from few-shot to many-shot model. It captures the implicit data augmentation on the model-level to improve the representation learning of tail items. 2. The item-level transfer connects head and tail items through item-level features, to ensure a smooth transfer of meta-mapping from head items to tail items. The two types of transfers are incorporated to ensure the learned knowledge from head items can be well applied for tail item representation learning in the long-tail distribution settings. Through extensive experiments on two benchmark datasets, results show that our proposed dual transfer learning framework significantly outperforms other state-of-the-art methods for tail item recommendation in hit ratio and NDCG. It is also very encouraging that our framework further improves head items and overall performance on top of the gains on tail items.
\end{abstract}
\maketitle

\section{Introduction}

\begin{figure}[t!]
\centering  \tiny
\includegraphics[width=3.4in]{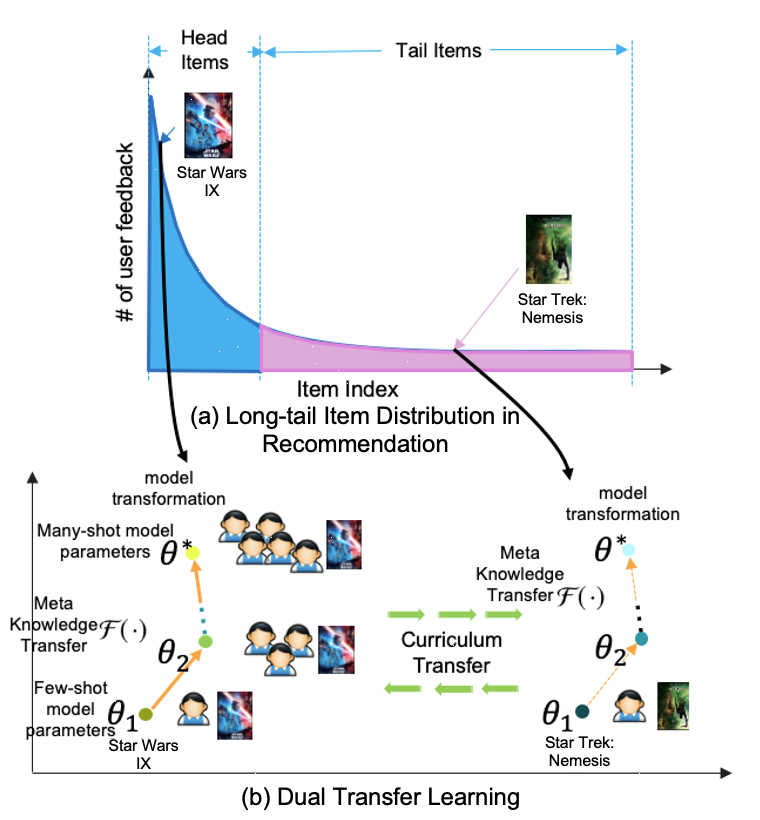}  
\vspace{-10pt}
 \caption{(a) An example of a highly-skewed long-tail distribution for items. \newline
 (b)
 The dual transfer learning framework: one is the meta-level knowledge transfer that learns model parameter changes from few-shot model (recommender trained based on few user feedback) to many-shot model (recommender trained based on rich user feedback), as shown in the orange arrow from $\theta_1$ to $\theta^*$; the other is curriculum transfer, shown in green arrow, that learns feature connections between head and tail items, such as the semantic relations between movie `Star Wars IX' and movie `Star Trek: Nemesis'. }
  \label{fig:intro} 
\end{figure}

Recommendation systems help users discover things they might be interested in, and have been widely adopted in various online systems, including e-commerce platforms (e.g., \cite{greenstein2018personal}, \cite{linden2003amazon}), video-sharing and video streaming websites (\cite{yi2019sampling}, \cite{hallinan2016recommended}), and online social platforms (\cite{abel2011analyzing}, \cite{pal2020pinnersage}). Building recommendation systems is challenging for many reasons. One of the most notorious problems is the prevalent long-tail distribution in user-item interactions \cite{liu2020long}: for items, a small fraction of popular items receive most of the user feedback, while most items only have few user feedback. As an example is illustrated in Figure \ref{fig:intro}(a), in Netflix, the movie of ``Star Wars IX'' contains many user feedback while users give less feedback for movie ``Star Trek: Nemesis''. As a result, recommendation models trained on dataset with such long-tail item distribution would easily overfit on a small fraction of popular items, and produce sub-optimal performance for the rest of the large chunk of tail items. Deploying such models in real world applications would also lead to the popularity bias \cite{abdollahpouri2017controlling} and ``rich gets richer'' feedback loop \cite{ma2020off}. Therefore, it becomes critical to address the long-tail item distribution problem in recommendations, especially to improve the recommendation of tail items.



A number of recent studies have aimed to address this problem, including resampling \cite{he2009learning} to rebalance the dataset, and reweighting \cite{cui2019class} to assign higher costs to tail items in the loss function. Besides these works, in recommendation, the tail items are usually treated as the cold-start items \footnote{In the work, cold-start items and tail items are both referred to the same group of items that contain few user feedback.}. Those recommenders usually leverage various content information (e.g. item title and cross-domain information) to enhance the representation learning of tail items. However, most existing recommenders focus on the tail items alone, without considering the connection with head items (popular items) -- which contain rich user feedback information and transferable contextual information related to tail items (e.g. belonging in the same genres).

In this work, we propose to transfer knowledge from head items that contain rich user feedback information to tail items that have few user feedback, to enhance the recommendation for tail items. Our evaluation criteria is that we expect to improve tail item recommendations while not hurting overall performance. Transferring the knowledge from head to tail items is challenging and non-trivial. Specifically, it poses \textbf{two key challenges}:
\squishlist
\item First, many existing transfer learning methods are pre-trained on a source task, and then fine-tuned on the final target task. The underlying assumption is that the similar data distribution between source task and target task ensures the high quality of knowledge transfer. This assumption does not suit well for long-tail setting where there is a large gap between the head and tail item distributions \cite{wang2017learning}. Hence, an important question is how to distill useful knowledge, that can be well transferred to tail items in the context of the imbalanced distribution.
\item Second, recommenders usually deal with a large catalog of various items. Head and tail items could show significant diversity in both user preference and content features. Thus another key challenge is how to avoid compromising overall model performance while improving on tail slices.
\squishend

With these challenges in mind, we propose a novel dual transfer learning recommendation MIRec that contains both (1) \underline{M}odel-level transfer (transfer learning across models) and (2) \underline{I}tem-level transfer (transfer learning across items), as shown in Figure \ref{fig:intro} (b). For (1), it learns a generic meta-level knowledge (meta-mapping) on the model-level, i.e., how the model parameters would change when more user feedback is added, highlighted in the orange arrows in Figure \ref{fig:intro}. As indicated by \cite{wang2017learning}, the learned meta-mapping can capture the implicit data augmentation in model-level without changing data distribution -- for example, given an item with a user feedback, the meta-mapping learns to implicitly add similar user who could give feedback to the item. Therefore, the learned meta-mapping can be leveraged to enhance the representation learning of tail items that contains few user feedback. For (2), it considers the divergence between head and tail items. We then utilize the curriculum transfer learning to connect head and tail items. The item-level transfer ensures the learned meta-mapping from head items can be suitable applied to tail items.

The \textbf{contributions} of this work are three-fold:

\squishlist


\item We propose a novel dual transfer learning framework that collaboratively learns the knowledge transfer from both model-level and item-level. Both components ensure the knowledge can be well utilized from head to tail items in the long-tail distribution settings.



\item We design a new curriculum that is capable to smoothly transfer the meta-mapping learned from head items to the long-tail items.

\item Extensive experimental results on two benchmark datasets show that MIRec consistently improves tail item recommendations significantly, while achieving solid improvements for the overall performance and head item performance for both hit ratio (HR@K) and NDCG@K. We also find the learned representation of tail items by MIRec preserve better semantics where similar items are close to each other in the embedding space.

\squishend


\section{Related Work}
\noindent \textbf{Long-tail Distribution}. The long-tail distributions are presented in many real-world datasets, which heavily influences task performance in different areas, such as image classification \cite{cui2019class}, natural language processing \cite{czarnowska2019don}, and recommendation \cite{liu2020long}. A commonly used strategy for the long-tail distribution problem is re-sampling to rebalance the dataset, including over-sampling and under-sampling  \cite{brownlee2020imbalanced, chawla2002smote, he2008adasyn,he2009learning}. The over-sampling that adds redundant samples from minor classes easily causes overfit to minor classes, whereas the under-sampling that eliminates the examples from the majority class could lose important samples, leading to sub-optimal performance. Another way to deal with the long-tail distribution is to refine the loss function, such as adding different weights/regularisations for classes/items \cite{cui2019class, beutel2017beyond}. For example, the recent proposed logQ corrections \cite{menon2020long,yi2019sampling,bengio2008adaptive} construct an item frequency-aware regularizor to alleviate the imbalanced influence. It is also worth to mention that there are some works explore other ways of dealing with the data long-tail distribution, such as meta-learning \cite{wang2017learning}, decoupling the learning process to only adjust classifier \cite{kang2019decoupling}.



In recommendations, the long-tail distribution of user feedback is more obvious, especially in item-side due to the rapid increase of large amount items. It heavily influences the recommendation performance \cite{beutel2017beyond,yin2012challenging,liu2020long,domingues2013combining}, especially for tail items. Some studies consider the long-tail item  distribution to improve the tail item recommendation, such as adding different weights to user-item pairs \cite{yin2012challenging} and clustering \cite{park2008long}. Another close related works are cold-start recommendations \cite{volkovs2017dropoutnet} that focus on the tail items, with less emphasize on the long-tail distribution. They usually integrate different kinds of user and item side information \cite{volkovs2017dropoutnet,zhu2019addressing,liang2020joint}, as well as incorporate different learning methods, such as meta-learning \cite{lee2019melu,dong2020mamo} and active learning \cite{zhu2019addressing}, to enhance the representation learning.




Different from them, in this work, we focus on transferring knowledge from head items that contain rich user feedback information to help learn tail items in the long-tail distribution, rather than purely based on cold-start items. 

\smallskip
\noindent \textbf{Meta-learning in Recommendation}. The meta-learning, known as the learning to learn, has attracted many attention due to its powerful performance in many applications. It aims to learn a model that captures the knowledge which can easily adapt to new task. Generally, there are three types of meta-learning approaches: 1. metric-based; 2. model-based; and 3. optimization based. Metric-based approaches \cite{snell2017prototypical} learn the distance across different tasks to uncover the general prototype. Model-based approaches \cite{santoro2016meta} focus on the model design to fast learn with few samples. Optimization-based approaches, like the model-agnostic meta-learning (MAML) \cite{finn2017model}, learn the general knowledge through the optimization among different tasks. Due to the meta-learning success for few-shot learning, recently, there are some attempts that explore the meta-learning framework for cold-start recommendation, such as utilizing the MAML to learn user different optimal parameters \cite{lee2019melu}, incorporating heterogeneous information network by item content information \cite{lu2020meta}, and memory-based mechanism  \cite{dong2020mamo}.


However, most of those methods focus on utilizing optimization to learn a generalized model (e.g. based on MAML) for cold-start recommendation, which does not fully utilize the long-tail distribution patterns (e.g. with less consideration about the distribution gap between head and tail items in the long-tail item distribution). In this work, we provide another meta-learning method that learns the meta-mapping from few-shot model (trained on small dataset) to many-shot model (trained on large dataset), to transfer the knowledge from head items to tail items. Perhaps the closest work to ours is Wang \etal \cite{wang2017learning} that investigated the image classification with long-tailed distribution dataset. It contains two key learners: base-learner and meta-learner. The base-learner is used to learn the image classification task, where the input is the sample features and output is the class label. The meta-learner learns a meta-level network that capture the model parameter changes from the few-shot model to many-shot model. That is, the meta-learner uses the few-shot model parameters as the input, to map/regress the parameters from many-shot model. Different from it, we adopt the meta-learning in recommendation, where there are substantially large amount of tail items and we further enhanced the meta-mapping to tail items by learning connections among items via curriculum learning.


\smallskip
\noindent \textbf{Curriculum Learning}. The curriculum learning, proposed by \cite{bengio2009curriculum}, has shown good performance in different areas \cite{guo2020breaking,el2020student}. The basic idea is to simulate the human learning process that humans can learn better if the examples are well organized (e.g. knowledge from easy to hard), and apply it to machine learning process. One important question is how to organize samples in the training process rather ran randomly sample, which is known as the curriculum. Recent studies have provided several curriculums and gained a good success, such as the teacher-student mechanics \cite{matiisen2019teacher}, the dynamic curriculum learning \cite{wang2019dynamic}. \cite{weinshall2018curriculum} also theoretically showed the curriculum learning can give a robust performance under different conditions. However, few work investigates the curriculum learning in recommendation, which we find especially useful when the long-tail distribution is considered to effectively transfer knowledge among head and tail items.



\section{Dual Transfer Learning Framework: MIRec}

\smallskip
\noindent \textbf{Problem Statement}. Suppose we have a set of users $\mathcal{U}$ and items $\mathcal{I}$, each user $u \in \mathcal{U}$ is represented by the feature vector $\mathbf{x}_u$ with the user ID and various user demographic information. Similarly for items, each $i \in \mathcal{I}$ is represented by the feature vector $\mathbf{y}_i$ with item ID and item content information. Furthermore, frequency of items from $\mathcal{I}$ follow a long-tail distribution, that a small portion of items in $\mathcal{I}$ received most of the feedback from users $\mathcal{U}$, as an example shown in Figure \ref{fig:intro} (a). Our goal is to improve recommendation quality on tail slices of the items, while keeping overall performance flat or better. We refer the task as \textit{long-tail recommendation} in the paper. Key notations are summarized in Table \ref{table:notation}.  


\smallskip
\noindent \textbf{Approach}. Considering the long-tail item distribution, inspired by \cite{wang2017learning}, \textit{we propose to transfer knowledge from items that contain rich user feedback (popular items) to items that have few user feedback (tail items)}, for the enhanced long-tail recommendation. Specifically, we propose a dual transfer learning framework that utilizes the long-tail distribution patterns to exploit both meta-knowledge in model level and feature connections in item level for enhanced transfer learning from head to tail:

\squishlist
\item Transfer learning \textit{across models}: it learns a meta-mapping from a few-shot model's (model trained on a small datasets \cite{ravi2016optimization}) parameters to a many-shot model's (model trained on a large dataset \cite{gui2018few}) parameters via a meta-learner $\mathcal{F}(\cdot)$. It captures a generic model meta-level knowledge of model transformation -- how model parameters are likely to change with more training data added, to help the learning of tail item that only have few user feedback \cite{gui2018few,wang2017learning}, as shown in Figure \ref{fig:intro}(b) orange arrows;

\item Transfer learning \textit{across items}: it connects head and tail items by their features to smoothly transfer the learned meta-level knowledge between head and tail items. To do so, we design a curriculum that re-organizes the training samples in different curriculum stages to transfer the learned features among head and tail items. The feature connection is used to ensure the learned meta-level knowledge can be applied for the tail items, as shown in Figure \ref{fig:intro}(b) green arrows.
\squishend




\begin{table}[t!]
      \caption{Notation.}
      \vspace{-10pt}
       \label{table:notation}
 \begin{center}
    \begin{tabular}{c|l} \toprule
     \textbf{Notation}     &  \textbf{Explanation}   \\
     \hline
     $\mathcal{U}$, $\mathcal{I}$ & user set, item set \\
     $\mathbf{x}_u$, $\mathbf{y}_i$ & user $u$ (item $i$) feature vector \\
     $I_h(k)$ & head item set of items that have no less \\ &than $k$ samples\\
     $I_t(k)$ & tail item set of items that have less than $k$ samples \\\hline
     $\Omega^*$ & training dataset for many-shot model\\
     $\Omega(k)$ & training dataset for few-shot model\\ \hline
     $g(\cdot)$ & base-learner\\
     $\mathcal{F}(\cdot)$ & meta-learner\\
     \bottomrule
   \end{tabular}
   \end{center}
\vspace{-10pt}
\end{table}





\subsection{Transfer Learning Across Models: Meta-level Knowledge Transfer} \label{sec:meta}

\begin{figure*}[t!]
\centering  \tiny
\vspace{-5pt}
\includegraphics[width=7.0in]{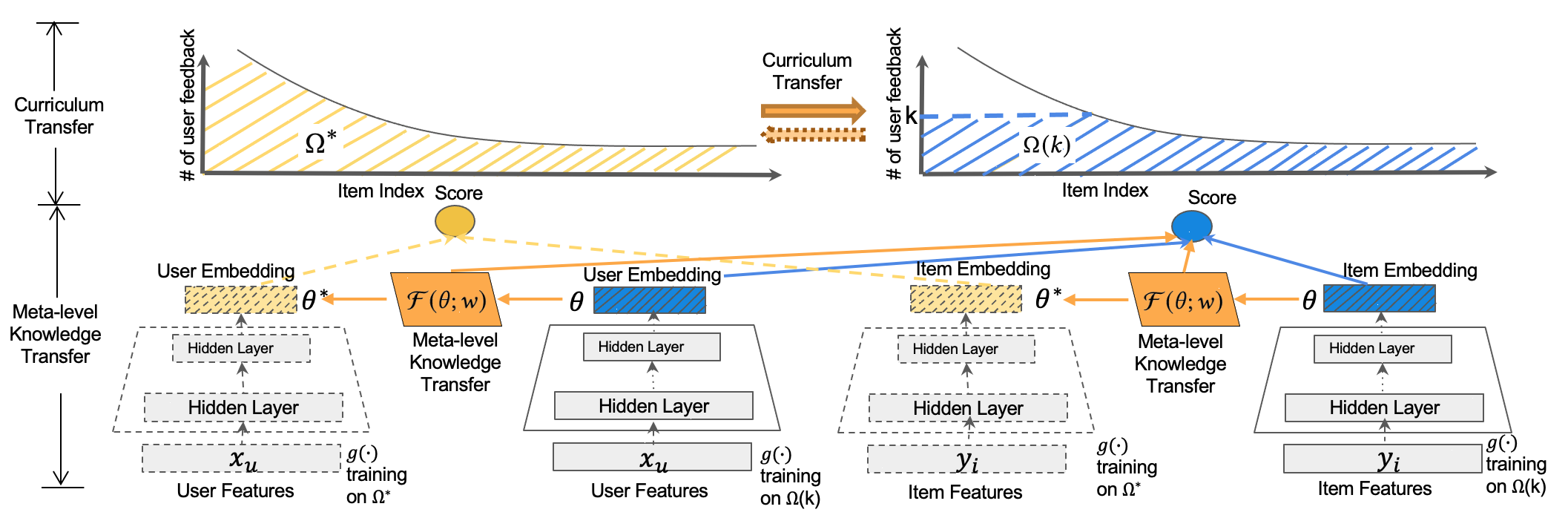}  
\vspace{-10pt}
 \caption{Framework of MIRec. It contains two types of transfers: (i) via the meta-learner $\mathcal{F}(\theta; w)$, the meta-level knowledge transfer learns the model parameter evolution from few-shot model parameters $\theta$ to many-shot model parameters $\theta^*$; (2) the curriculum transfer learns the connections between head and tail items to smooth transfer the meta-level knowledge from head to tail. }
 \label{fig:frame} 
\vspace{-5pt}
\end{figure*}







The recommender performance usually suffers for tail items. As indicated by \cite{yin2012challenging,cui2018large}, one of the main reason is lack of data points \cite{yin2012challenging}, and thus cause a weak item representation (tail items are underrepresented). On the other hand, we have fairly rich information and sufficient data points for items on the popular side. Inspired by meta-learning \cite{wang2017learning}, we propose to explore the connection (i.e. meta-learner) between models learned with only a few training examples, and the models learned from the same item but with sufficient training examples through model parameters. Similar as \cite{wang2017learning}, the assumption is that the meta-learner can capture the implicit \textit{data augmentation} \cite{wang2017learning}. For example, given an item with a user feedback, the meta-learner learns to implicitly add similar users who could have provided feedback to the same item. But rather than explicitly adding similar data point, the meta-learner learns the impact on the learned model parameters -- refer as meta-level knowledge. It essentially provides a ``magic wand'' for us to improve representation quality even when we do not have sufficient data.


In this section, we introduce the meta-level knowledge transfer in our framework, as illustrated in the bottom part in Figure \ref{fig:frame}. There are two key learners in the meta-level knowledge transfer: (i) \textbf{base-learner} $g(\mathbf{x}_u, \mathbf{y}_i; \theta)$: It learns the user preference towards items. It takes a pair of feature vectors $\mathbf{x}_u$ and $\mathbf{y}_i$ for user $u$ and item $i$ as the inputs, with model parameters $\theta$. The output is the user $u$'s preference score towards item $i$. By using different training data, $g(\mathbf{x}_u, \mathbf{y}_i; \theta)$ is applied for both few-shot model learning and many-shot model learning. (ii) \textbf{meta-learner} $\mathcal{F}(\theta; w)$: It learns the meta-mapping from few-shot model parameters to many-shot model parameters. Its input is model parameters $\theta$ in $g(\mathbf{x}_u, \mathbf{y}_i; \theta)$ which is trained by $\Omega(k)$, with meta-learner model parameter $w$.


\smallskip
\noindent \textbf{Base-Learner}: With the user and item feature vectors $\mathbf{x}_u$ $\mathbf{y}_i$, the user preference estimation by base-learner $g(\cdot)$ is conducted by a two-tower neural network architecture. It can well encode various types of features for users and items, and is scalable to large corpus dataset \cite{yi2019sampling,ma2020off}.




Concretely, as an example shown in Figure \ref{fig:frame} in light yellow, the two-tower neural network first learns latent representations (embeddings) given user and item features, through multilayer perceptron (MLP) models (referred as the towers). We denote the user tower as $g_u(\mathbf{x}_u; \theta)$ and item tower as $g_i(\mathbf{y}_i; \theta)$. Then, by combining the user and item embedding through inner product, the output is the user preference towards the item $s(\mathbf{x}_u, \mathbf{y}_i; \theta) = <g_u(\mathbf{x}_u; \theta), g_i(\mathbf{y}_i; \theta)>
$.

We formalize the recommendation task as a multi-class classification problem. We use softmax as the loss function to learn the probabilistic distribution of user $u$'s preference towards different items for recommendation, as defined below:


\begin{equation}
    p(\mathbf{y}_i|\mathbf{x}_u; \theta) = \frac{e^{s(\mathbf{x}_u, \mathbf{y}_i; \theta)}}{\sum_{j \in \mathcal{I}} e^{s(\mathbf{x}_u,\mathbf{y}_j; \theta)}}.
\label{eq:softmax}
\end{equation}
Therefore, the loss function for base-learner $g(\cdot)$ can be formulated as:
\begin{equation}
L_g(\theta | \Omega) = -\frac{1}{|\Omega|}\sum_{(u,i, r(u, i)) \in \Omega} r(u, i)logp(\mathbf{y}_i|\mathbf{x}_u; \theta),
\label{eq:lg}
\end{equation}
where the $r(u, i)$ is the reward function and  $\Omega$ is the training dataset.

For $r(u, i)$, a binary reward can be defined as:
\begin{equation}
r(u, i) =
  \begin{cases}
    1       & \quad \text{if user $u$ gives positive feedback, e.g. click, on item $i$},\\
    0  & \quad \text{otherwise}.
  \end{cases}
\end{equation}
By adding the rewards in the loss function, it would learn high preference score for items that user engaged with. Other rewards like the continuous rewards of user engagement (e.g., video watch time), can also be easily integrated. In the following, we focus on the binary rewards for illustration.

For training data $\Omega$, it is generated by a set of triplets $(u, i, r(u, i))$ that contains a user $u$, item $i$ and their rewards $r(u, i)$. For example, 
\begin{equation*}
\Omega := \{(u, i, r(u, i))\}. 
\end{equation*}
We can use different $(u, i, r(u, i))$ to construct the the training dataset. By turning $g(\cdot)$ on those training dataset with Equation \eqref{eq:lg}, we can get few-shot models (models are trained with few data samples, such as down sample user feedback for each items) and many-shot models (models are trained with rich data samples, such as use all the user feedback). We denote the training dataset that is used to train the few-shot (many-shot) model as $\Omega(k)$ ($\Omega^*$). Details of generating $\Omega(k)$ and $\Omega^*$ in MIRec are discussed in section \ref{sec:curr}.

\smallskip
\noindent \textbf{Meta-Learner}. With the learned few-shot model and many-shot model by base-learner $g(\cdot)$, we introduce the meta-learner $\mathcal{F}(\cdot)$ that maps the two types of model parameters, to capture the meta-level knowledge -- the evolution of model parameter changes when more training examples are added. To do so, we first learn the many-shot model parameters $\theta^*$. Then we present the final objective function that simultaneously trains the few-shot model and meta-mapping.

To learn $\theta^*$, we optimize the $g(\cdot)$ by using $\Omega^*$ as the training set:
\begin{equation}
\theta^* = \argmin L_g(\theta | \Omega^*).
\label{eq:lossg}
\end{equation}
where $L_g(\cdot)$ is defined in Equation \eqref{eq:lg}.


With $\theta^*$ and $\Omega(k)$, then the final objective function is:
\begin{align}
\begin{split}
L(w, \theta|\Omega^*, \Omega(k)) &= ||\mathcal{F}(\theta; w) - \theta^*||^2 + \lambda L_g(\theta | \Omega(k)),
 \label{eq:loss}
 \end{split}
\end{align}
where $||\cdot||$ is the l2 norm. $\lambda$ is the regularization parameter balance the two terms. For each term in Equation \eqref{eq:loss}: (i) $||\mathcal{F}(\theta; w) - \theta^*||^2$ minimizes the distance between the predicted many-shot model parameters (by $\mathcal{F}(\theta; w)$) and many-shot model parameter $\theta^*$. It ensures the predicted many-shot model parameters learned by meta-mapping can well fitted for the data samples in $\Omega^*$; (ii) the second term $L_g(\theta | \Omega(k))$ trains the base-learner based on training dataset $\Omega(k)$, to encourage the high recommendation performance for data samplings in $\Omega(k)$. The $L_g(\cdot)$ is the same as Equation \eqref{eq:lg}. A good performance of base-learner could help the learning of meta-mapping that also maintained high accuracy on the base recommendation task. Therefore, the loss function that simultaneously trains the few-shot model and meta-mapping offers both the high-quality of base-learner recommendation and high accuracy of meta-learner model mapping. 

\textbf{Implementations of $\mathcal{F}(\cdot)$}. In practice, the base-learner $g(\cdot)$ could be very complex, such as the two-tower model where each tower could contain many layers. In this situation, $g(\cdot)$ has a large number of model parameters. Directly mapping all those $g(\cdot)$ model parameters by $\mathcal{F}(\cdot)$ could be both high computational cost and unstable \cite{wang2016learning,andrychowicz2016learning}. Therefore, similar as \cite{wang2017learning}, we focus on the model parameters in the last layer of user and item towers for meta-learning. In this way, the meta-learning is capable to be more efficient and generalized to learn different items, which is especially helpful for the large number of various tail items in recommendation. For $\mathcal{F}(\cdot)$, different networks can be applied, such as regression networks \cite{wang2016learning}, knowledge graph \cite{peng2019few}, and DNN-based network \cite{sun2019meta}. Here we use a fully connected layer as the mapping network for $\mathcal{F}(\cdot)$ , which we empirically find is a relative simple and effective approach that can gain good performance.

\subsection{Transfer Learning Across Items: Curriculum Transfer}  \label{sec:curr}
As shown in Section \ref{sec:meta}, both the training dataset $\Omega(k)$ for few-shot model and $\Omega^*$ for many-shot model play a pivotal role to ensure the high quality of learned model generic transformation by $\mathcal{F}(\cdot)$. 

A traditional way to construct the $\Omega(k)$ and $\Omega^*$ is only based on the head items. For example, \cite{wang2017learning} selects items that have no less than k user feedback, denoted as $I_h(k)$, and uses all the user feedback from item $i \in I_h(k)$ to construct the training dataset for many-shot model. The few-shot model is learned by randomly sampling a smaller fixed number of user feedback to $i \in$ $I_h(k)$. Therefore, the meta-knowledge of model parameter evolution is only based on the head items. Since the learned meta-mapping is directly applied on the few-shot models of tail items $i \in I_t(k)$ ($I_t(k)$ is the set of items that have less than k user feedback), a basic assumption for the directly mapping is that the meta-mapping for head and tail items are the same.

However, different from many other classification problem (e.g. image classification), the number of items (i.e., classes) for recommenders is highly diverse. In addition, there are way more tail items with few observations than the head items. In this situation, the meta-mapping that are learned only based on the head items would constrain the mapping of the large amount of tail item representations, leading to sub-optimal performance. Thus, we introduce curriculum learning to smoothly transfer the meta-mapping from head items to tail items by construction $\Omega^*$ and $\Omega(k)$.

Concretely, we propose a Long-Tail distribution-aware CURriculum strategy (LTCur) that can both well transfer the learned features from head items to tail items, and at the same time, relatively keeping the original distribution to alleviate the influence of distribution bias. Therefore, we construct the two sets $\Omega^*$ and $\Omega(k)$ to ensure both head and tail items are trained in the training stages.

In LTCur, the training set $\Omega^*$ for many-shot model is defined as:
\begin{equation}
    \Omega^*:= \{(u, i, r(u, i))\}
\label{eq:setstar}
\end{equation} 
that $\forall (u, i, r(u, i)) \in \Omega^*$ satisfies:
\squishlist
\item $u\in \mathcal{U}, i \in \mathcal{I}$;
\item $\sum_{(u, i, r(u, i)) \in \Omega^*}r(u, i) = \sum_{u \in U}r(u, i)$, for $i \in \mathcal{I}$;
\squishend
That is, as shown in Figure \ref{fig:frame} of item distribution in yellow part, $\Omega^*$ includes the all the user feedback from both the head items and tail items. To train $g(\cdot)$ on $\Omega^*$, considering the large number of tail items, we add the logQ correction \cite{yi2019sampling}:
\begin{equation}
s^c(\mathbf{x}_u, \mathbf{y}_i; \theta^*) = s(\mathbf{x}_u, \mathbf{y}_i; \theta^*) + \lambda_c\ log(p_i),
\label{eq:logq}
\end{equation} 
where the $\lambda_c$ is the weight for the correction. $p_i$ is the sampling probability of item $i$, which is usually calculated by the frequency of item $i$ divided the total number of training samples.

For the training set $\Omega(k)$ that is used to train the few-shot model parameters, as shown in Equation \eqref{eq:loss}, we construct it as:
\begin{equation}
    \Omega(k):= \{(u, i, r(u, i))\}
\label{eq:setkf}
\end{equation} 
that $\forall (u, i, r(u, i)) \in \Omega(k)$ satisfies:

\squishlist
\item $u\in \mathcal{U}, i \in \mathcal{I}$;
\item $\sum_{(u, i, r(u, i)) \in \Omega(k)}r(u, i) = k$, if $i \in I_h(k)$;
\item $\sum_{(u, i, r(u, i)) \in \Omega(k)}r(u, i) = \sum_{u \in U}r(u, i)$, if $i \in I_t(k)$;
\squishend
that is, we consider all the user feedback for the tail items. The new constructed  training set $\Omega^*$ and $ \Omega(k)$ ensure (1) tail items are fully trained in both the many-shot model and few-shot model to ensure the high quality of the learned item representations in both many-shot and few-shot models. It can further act as the cornerstone for the high performance of meta-mapping. (2) In the few-shot model training, the distribution of tail items relatively keeps the same as the original distribution. It can alleviate the bias among tail items that brings by the new distribution.  

The two training datasets $\Omega^*$ and $ \Omega(k)$ both contain the head and tail items but with different head and tail distributions. 
Thus, based on the curriculum learning \cite{bengio2009curriculum,cui2018large}, the meta-mapping $\mathcal{F}(\cdot)$ not only learns the mapping from few-shot model parameters to many-shot model parameters, but also feature connections between head and tail items through the curriculum. The dual transfer framework ensures the smooth meta-mapping transfer from head to tail items. Experiments verify that our proposed curriculum can improve the performance of both overall and tail item recommendation performance. Details of the implementation of MIRec is shown in Algorithm \ref{alg:MIRec}.

\begin{algorithm}[t]
    \SetKwInOut{Input}{Input}
    \SetKwInOut{Output}{Output}
    \SetKwInOut{Function}{Function}
    \Input{user set $\mathcal{U}$, item set $\mathcal{I}$, feature vector $\mathbf{x}$, $\mathbf{y}$ that maps each user and item input to an embedding spase, step size $\alpha$, $\beta$, $\gamma$, batch size $B$}
    \BlankLine
    Initialized Model parameters $\theta$, $w$\;
    Construct sets $\Omega^*$, $\Omega(k)$ based on Equation \eqref{eq:setstar} and \eqref{eq:setkf};

\SetKwFunction{OP}{OPTIMIZE\_PARAMETERS}
    \SetKwProg{Fn}{Function}{:}{}
    \Fn{\OP{$\Omega$, $g(\cdot)$}}{
    \# Calculate the optimize model parameters for many-shot model\;
    randomly initialize $\theta^*$\;
    \While{not converged}{
    Sample batch of user and item pairs $\{(u, i, r(u, i))\}_{m = 1}^B \in \Omega$\\
      \For{each $(u, i, r(u, i))$}{
      $\theta^* \leftarrow \theta^* - \alpha \bigtriangledown_{\theta^*} L_g(\theta^*|\Omega)$
      } 
      
     }
 \textbf{return} $\theta^*$\; 

}

$\theta^* \leftarrow$ \OP{$\Omega^*$, $g(\cdot)$}\;

\While{not converged}{
Sample batch of user and item pairs $\{(u, i, r(u, i))\}_{m = 1}^B \in \Omega(k)$\\
\For{each $(u, i, r(u, i))$}{
  \For{local update steps}{
  $\theta \leftarrow \mathcal{F}(\theta; w)$\;
  calculate  $L_g(\theta | \Omega(k))$\;
  $\theta \leftarrow \theta - \beta \bigtriangledown_{\theta} L(w, \theta|\Omega^*, \Omega(k))$
  } 
  Global update both meta-learner model parameters $w$ in $\mathcal{F}(\cdot)$ and few-shot model parameter $\theta$ based on Equation \eqref{eq:loss}:\
  $w \leftarrow w - \gamma \bigtriangledown_wL(w, \theta|\Omega^*, \Omega(k))$\;
  $\theta \leftarrow \theta - \gamma \bigtriangledown_{\theta}L(w, \theta|\Omega^*, \Omega(k))$\;
 }
 }
 \textbf{return} $\theta$, $w$
    
    \caption{Training procedure of MIRec.}
    \label{alg:MIRec}
\end{algorithm}

\subsection{Prediction by MIRec}
The meta-level knowledge transfer learns the model parameter evolution when more training data is available. Different from tail items, head items contain rich user feedback. since $\mathcal{F}(\cdot)$ is learned to match $\theta^*$ by Equation \eqref{eq:loss}, the base-learner could offer a better representations for head items rather than ones by meta-mapping. Therefore, to predict the user preference towards both head and tail items, we combine the representation learned from both the base-learner and meta-mapped base-learner \cite{wang2017learning}. One example to integrate their representations for user preference learning is:
\begin{align}
     s_{MIRec}(\mathbf{x}_u, \mathbf{y}_i) &= s(\mathbf{x}_u, \lambda_p*\mathbf{y}_i; \theta^*) + s(\mathbf{x}_u, (1 - \lambda_p)*\mathbf{y}_i; \mathcal{F}(\theta; w)) \nonumber\\
     &=\lambda_p g(\mathbf{x}_u,  \mathbf{y}_i; \theta^*) + (1 - \lambda_p)g(\mathbf{x}_u,  \mathbf{y}_i; \mathcal{F}(\theta; w)),
\label{eq:pred}
\end{align}
where $\lambda_p$ is used to balance the meta-mapping based model and many-shot based model. And we experimentally find it has a good performance.

\section{Discussion}

\textbf{Expansion of MIRec}. In this work, we construct two subsets $\Omega^*$ and $\Omega(k)$ for transfer learning under the long-tail distribution, where $k$ is a fixed constant. To make the transfer learning more smooth among items, we can further vary the $k$ to construct different sets and recursively learn the model dynamic trajectories. Concretely, we can select a sequence of $\{k_j\}$ that satisfies $k_1 < k_2 < k_3, ... < k_m$. Then the corresponding $\Omega(k_j)$ is constructed based on Equation \eqref{eq:setkf}. The meta-learner would be a sequence of $\mathcal{F}_{1, 2}(\cdot), ... \mathcal{F}_{j, j+1}(\cdot) ... \mathcal{F}_{m-1, m}(\cdot)$, where $\mathcal{F}_{j, j+1}(\cdot)$ that learned the model mapping from $\Omega(k_j)$ to $\Omega(k_{j + 1})$ in a recursive way \cite{wang2017learning}. And the final prediction of user $u$ preference towards item $i$ can be calculated by:

\begin{align*}
\begin{split}
   \lambda_{p, 0}g(\mathbf{x}_u,  \mathbf{y}_i; \theta^*) + &\sum_{j = 1}^{m-1} \lambda_{p, j}g(\mathbf{x}_u,  \mathbf{y}_i; \mathcal{F}_{j, j+1}(\theta_{j, j+1}; w_{j, j+1})), \\ &where \ \sum_{j = 0}^{m-1} \lambda_{p, j}= 1.
\end{split}
\end{align*}

\smallskip
\noindent \textbf{Long-tail distribution from user side}. So far we focus on the long-tail distribution from item side. In practice, MIRec can also be adopted when we focus on the long-tail distribution from user side or from both user and item side. When only the user side long-tail distribution is considered, we can simply change the MIRec by reversing the user $u$ and item $i$. And the $\Omega^*$ and $\Omega(k)$ is constructed based on the user distribution, correspondingly. When the long-tail distribution is considered from both user and item sides, one simple way that we can do is to use both user- and item-based MIRec, and integrate their results as many ensemble methods.

\section{Experiment}
In this section, we evaluate MIRec for recommendation tasks through experiments on two datasets. We seek to address the following key research questions:




\textbf{RQ1}: How well does the dual transfer learning framework MIRec perform compared to the state-of-the-art methods?

\textbf{RQ2}: How do different components (meta-learning and curriculum learning) of MIRec perform individually? Could they complement each other?


\textbf{RQ3}: How does our proposed curriculum learning strategy compare with the alternatives?

\textbf{RQ4}: Besides downstream task performance, are we actually learning better representations for tail items? Could we see the differences visually?

\subsection{Datasets}

We adopt two widely used benchmark datasets: MovieLens 1M \footnote{https://grouplens.org/datasets/movielens/1m/} and Bookcorsssing \footnote{http://www2.informatik.uni-freiburg.de/\textasciitilde cziegler/BX/}, both with rich user and item information.
We follow the similar procedures adopted in \cite{kang2020learning,he2017neural,lee2019melu,tang2018personalized} to engineer the features and labels. Specifically, all the <user, item> pairs with an explicit ratings are considered as positive examples ($1$s), and the rest of the pairs are all considered as negative examples ($0$s). Among these, any pairs of <user, item> with invalid / missing features are filtered, as dealing with the missing value is often treated as a separate issue \cite{yin2012sparsity,nazabal2020handling}. Besides age and year which are treated as the continuous features, all the other user and item features are treated as categorical feature and represented by one-hot encoding. We apply bag-of-words representation for item title feature \cite{soumya2014text}. A summary of the two datasets is shown in Table \ref{table:stat}.



\begin{figure}[t!]
\centering  \tiny
\vspace{-5pt}
 \includegraphics[width=3.3in]{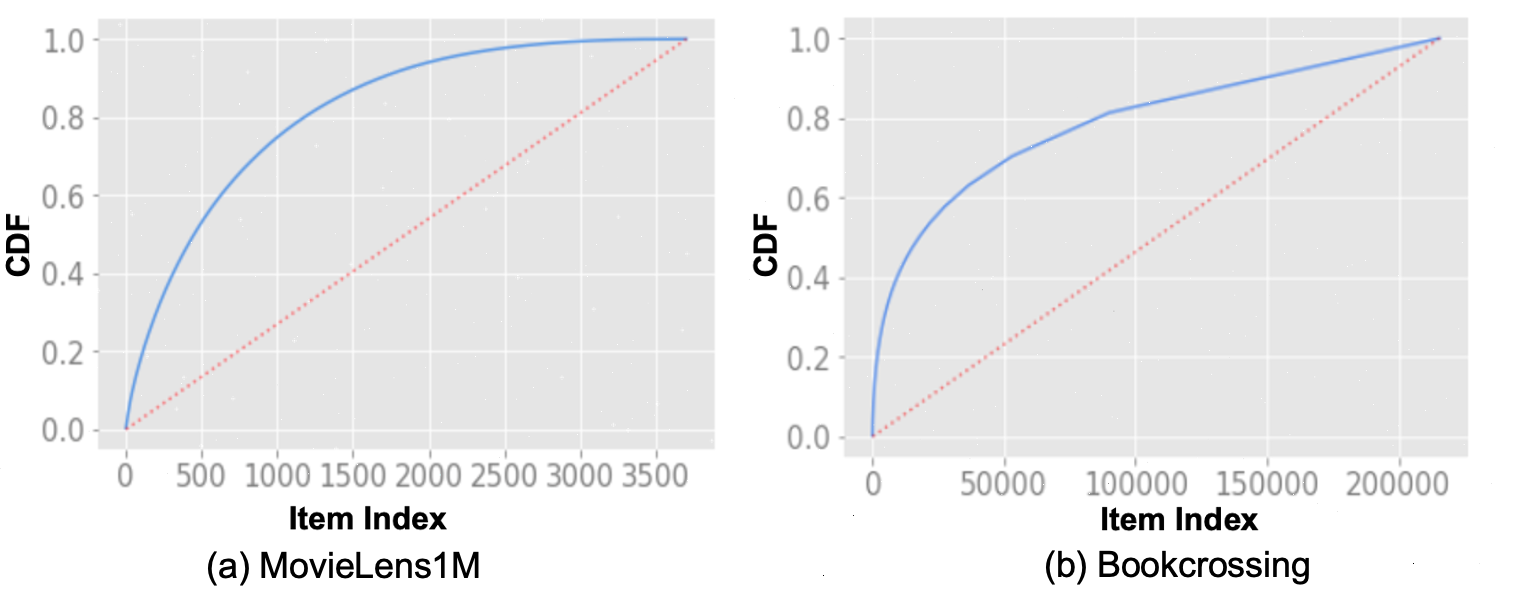}  
\vspace{-10pt}
 \caption{The CDF of the item distribution in MovienLens1M and Bookcorssing. It shows items in both datasets are in the long-tail distributions.}
  \label{fig:cdf} 
\vspace{-5pt}
\end{figure}

In both datasets, the item distribution follows a highly-skewed long-tail distribution (see Figure ~\ref{fig:cdf}). Especially for the bookcrossing dataset, a small portion of popular items contributes more than half of all the user feedback. It shows the two datasets are well-suited for our target problem of improving tail recommendation quality in the presence of long-tail item distribution.



Consistent with the prior work \cite{he2017neural,kang2020learning}, we divide the data into three parts by leave-one-out evaluation: for each user, the most recent item is for testing, the second most recent interacted item is for validation, and the rest of the items are for training. Since the Bookcrossing dataset does not have the timestamp information, for each user, we randomly select one item for testing, one for validation, and use the rest of items for training.


\begin{table}[t!]
\caption{Summary of the datasets and content information.}
\vspace{-5pt}
\label{table:stat}
\begin{tabular}{c|cc}\toprule 
& \textbf{MovieLens1M}       & \textbf{Bookcrossing}                   \\ \hline
\# User              & 6,040                                                                      & 50,454                      \\
\# Items             & 3,706                                                                      & 222,154                        \\
\# Feedback & 1,000,209                                                                  & 1,031,175                      \\
Sparsity             & 95.5316\%                                                                  &   99.9908\%                    \\ \hline
User Features         & \begin{tabular}[c]{@{}@{}l@{}}IDs, Gender, Occupation,\\ ZipCode, Age\end{tabular} & IDs, Location, Age                  \\ \hline
Item Features         & \begin{tabular}[c]{@{}l@{}}IDs, Title, \\ Genres, Year\end{tabular}             & \begin{tabular}[c]{@{}l@{}}IDs, Title, Author, \\ Year, Publisher\end{tabular} \\ \bottomrule
\end{tabular}
\vspace{-5pt}
\end{table}

\subsection{Setup}

\textbf{Evaluation Criteria}. 
We adopt two common evaluation metrics, Hit Ratio at top K (HR@K) and NDCG at top K (NDCG@K) \cite{he2017neural}. HR@K measures whether the test item is retrieved in the top K ranked items. The NDCG@K considers the position of correctly recommended items by giving higher scores to the top ranked items.

With the goal of improving the tail item recommendation quality without hurting the overall performance, we further report the metrics evaluated on the tail and head item set respectively. Based on the Pareto Principle \cite{box1986analysis} and the item distributions (Figure \ref{fig:cdf}), we split the first 20\% most frequent items in Movielens1M and 0.1\% items in Bookcrossing \footnote{The split rate of 0.1\% can better show the performance differences among methods, especially for tail items.} for head items, and the rest items are treated as tail items.


The two metrics above are usually used to evaluate the overall model performance, regardless of the data distribution. Our goal of the framework is to build a single model to improve the tail item recommendation without hurting the overall performance by considering the long-tail item distribution. Thus, we report model performance on head, and tail slices respectively, as well as the overall performance. We describe how we claim success for our recommenders in Table \ref{table:evalm}, with the assumption that we are building one single model to serve both head and tail recommendations. Alternatively, if we were to build separate models for head and tail slices, most likely we would be able to achieve better results from both models on their corresponding data slices.

As shown in the Table \ref{table:evalm}, we consider a recommender has `good results' when the model improves on the tail, while taking a hit on the head but still keeping overall performance flat. The results are even `better' when we could keep the performance on head items neutral while improving both on tail and overall. The best scenario would be achieving improvements across the board on head, tail, and overall.

\begin{table}[t!]
\caption{Evaluation criteria w.r.t. performance improvement on different data slices.}
\vspace{-10pt}
\label{table:evalm}
\begin{tabular}{c|c|c|c} \toprule 
 HR@K/NDCG@K                & \textbf{Overall} & \textbf{Head Items} & \textbf{Tail Items} \\ \hline
Good Results     &   --      &  $\searrow$          &    $\nearrow$        \\ \hline
Better Results   &   $\nearrow$      &   --         &     $\nearrow$       \\\hline
Great Results &    $\nearrow$     &    $\nearrow$        &    $\nearrow$       \\\bottomrule
\end{tabular}
\vspace{-10pt}
\end{table}

\smallskip
\noindent \textbf{Baselines}. To fully demonstrate the effectiveness of MIRec on long-tail item distribution, we compare MIRec with comparative model-agnostic methods, which includes strategies such as training distribution re-sampling, loss function refinement and various curriculum learning and meta-learning strategies. For fair comparison, similar as \cite{ma2020off}, we consider the same content-aware two-tower DNN model \cite{yi2019sampling} as the backbone model to establish a reasonable benchmark. The same number of layers and dimensions are used for all the strategies and MIRec for fair comparison. We list all the baseline models here: 


\smallskip
\noindent Backbone Model:
\squishlist
\item\textit{Two-tower Model} \cite{yi2019sampling, yang2020mixed, wang2019improving}: This model architecture has been proven to be highly effective in content-aware settings. It also demonstrates great scalability due to the efficiency of inner product for inference.
\squishend

\noindent Re-sampling Strategies:
\squishlist
\item \textit{Over-sampling} \cite{brownlee2020imbalanced}: A training data re-sampling strategy that repeatedly samples from tail items to re-balance the head items and tail items in the training distribution. The over-sampling strategy is more common in practice since user feedback data is highly valuable, and we do not want to down-sample on the positives.


\item \textit{Under-sampling} \cite{brownlee2020imbalanced}:
In contrast to over-sampling, we keep the tail items unchanged, and down-sample the head items. 

\squishend

\noindent Loss Function Refinement:
\squishlist
\item \textit{ClassBalance} \cite{cui2019class}: This is a recent state-of-the-art method for tackling the long-tail distribution in image classification settings. It first calculates the effective number of samples for each imbalanced class. Then a class balanced loss function is designed by using different weights of each class based on the effective number of samples. We adopt this approach method to the recommendation task, where items are treated as classes. 

\item \textit{LogQ} \cite{menon2020long}: The logQ correction is widely used to address the long-tail distribution problem in different areas. It corrects the predicted logits in the loss function by adding a correction term based on estimated item frequency.

\squishend

\noindent Curriculum Learning Strategies:
\squishlist
\item \textit{Head2Tail}: Based on the previous studies \cite{beutel2017beyond}, the recommendation performance on tail items is significantly worse than the head items, and the prediction variances of tail items are high. Thus, following the curriculum learning idea, we first train the two-towel model with head items in the first curriculum stage, then fine-tune the model with tail items only in the second curriculum stage.


\item \textit{Tail2Head}: We also explore the recommendation performance of Tail2Head which adopts the exact opposite of the curriculum of Head2Tail. That is, it first trains on the tail items, then fine-tunes the model on the head items. 
\squishend

\noindent Meta-learning Strategies:
\squishlist
\item \textit{MeLu} \cite{lee2019melu}: This is a recent state-of-the-art meta-learning based method for cold-start item rating prediction. It applied the optimization-based meta-learning (MAML) for rating prediction in the cold-start problem. We modify it to be applicable for the implicit feedback prediction task. For a fair comparison, we also applied the two-tower model as the backbone model for MeLu. 
\squishend


\smallskip
\noindent \textbf{Hyper-parameter settings}. We set the embedding dimension as 64 across all the different baselines and treatment models. For approaches that adopted the two-tower model architecture, we used exactly the same numbers of hidden layers, and hidden units. Concretely, we employ a widely used tower structure, where hidden dimensions of higher layers have 1/2 number of neurons of its next lower layer. The ReLU function is used as the active function. For methods that consider curriculum learning, we consider 100 training epochs for each curriculum stage (two stages in total) for a fair comparison. For non-curriculum-based methods, the total number of epochs is 200. The decay rate between each curriculum stages is determined by grid search from \{0.1, 0.01, 0.001, 0.0001\}. Similar as the other meta-learning \cite{lee2019melu}, we perform one gradient decent for local update in each step. The k in $\Omega(k)$ is determined by the split of head and item items. Regularization parameters are determined by grid search in the range of \{0.3, 0.1, 0.01, 0.001, 0.0001, 0.00001, 0.000001\}. Dropout rate and learning rate are determined by grid search in the range of \{0.1, 0.3, 0.5, 0.7, 0.9\} and \{0.1, 0.01, 0.001, 0.0001, 0.00001\} respectively. The batch size is set to 1024, to be consistent with previous results \cite{yi2019sampling}. The Adaptive moment estimation (Adam) is used to optimize MIRec. MIRec is implemented with TensorFlow Recommenders (TFRS) \footnote{https://blog.tensorflow.org/2020/09/introducing-tensorflow-recommenders.html}.




\begin{table*}[t!]
\caption{The recommendation performance of MIRec versus baselines on MovieLens1M. }
\vspace{-10pt}
\label{table:perfmovie}
\begin{tabular}{c|cc|cc|cc}
\toprule
\multirow{2}{*}{Measure\%}          & \multicolumn{2}{c|}{Overall} & \multicolumn{2}{c|}{Head} & \multicolumn{2}{c}{Tail} \\ \cline{2-7}
          & HR@10        & NDCG@10       & HR@10      & NDCG@10      & HR@10      & NDCG@10     \\\hline
Two-tower & 6.74         &  3.26             &   8.98         &  4.33            &   3.22         &    1.59         \\ \hline
Over-sampling & 0.23         &   0.10            &   0.14         &    0.04          &   0.37         &  0.19           \\ 
Under-sampling &  0.72        &  0.33             &  0.41          &    0.20          &   1.19        &  0.53           \\ \hline
ClassBalance      &   6.32           &     2.98          &  9.10          &      4.43        &   1.45        &  0.71 \\
LogQ      & 2.24             &  0.95             &    1.41        &      0.54        &      3.53      &  1.59           \\ \hline
Head2Tail &  1.31            &    0.66           &     1.90       &     0.97         &   0.40         & 0.18            \\
Tail2Head &  1.71            &  0.85            &   2.74         &   1.34           &     0.09       &  0.07           \\ \hline
MeLu &    5.96         &    2.88         &  8.98         &  4.35          &   1.23       &   0.58          \\ \hline
MIRec     &  \textbf{7.02}            &     \textbf{3.36}          &    \textbf{9.13}        &     \textbf{4.36}         &    \textbf{3.70}        &    \textbf{1.81}        \\ \bottomrule
\end{tabular}
\end{table*}

\begin{table*}[t!]
\caption{The recommendation performance of MIRec versus baselines on Bookcrossing.}
\vspace{-10pt}
\label{table:perfbx}
\begin{tabular}{c|cc|cc|cc}
\toprule
\multirow{2}{*}{Measure\%}           & \multicolumn{2}{c|}{Overall} & \multicolumn{2}{c|}{Head} & \multicolumn{2}{c}{Tail} \\ \cline{2-7}
          & HR@100        & NDCG@100       & HR@100     & NDCG@100      & HR@100      & NDCG@100     \\\hline
Two-tower &  4.98        &      1.56         &   54.67        &    17.58         &   0.32         &   0.05        \\ \hline
Over-sampling &  0.17       &    0.03           &  0.51          &      0.12        &     0.14       &     0.02        \\ 
Under-sampling & 0.76         &    0.16           &  3.56         &   0.73           &      0.51     &     0.11        \\ \hline
ClassBalance      &   4.58           &    1.06           &     49.62       &    11.77          &   0.45        &  0.08 \\
LogQ      &  2.06          &     0.49        &   15.52      &   4.11         &   0.82        &     0.15        \\ \hline
Head2Tail &    2.66          &      0.58         &  24.68         &  5.66           &    0.65      &   0.06        \\
Tail2Head  &   4.82           &      1.07         &   52.00        &     12.73         &   0.03         &    0.01            \\ \hline
MeLu &  3.33           &    0.75         &    30.28       &     7.22         &   0.86       &0.16             \\ \hline
MIRec     &  \textbf{5.25}           &     \textbf{1.60}          &    \textbf{55.28}      &   \textbf{17.29}          &    \textbf{0.58}      &    \textbf{0.14}      \\ \bottomrule
\end{tabular}
\vspace{-10pt}
\end{table*}

\subsection{Recommendation Performance (RQ1)}
We first compare methods for top-K recommendation on head, tail items and overall performance. Then we vary the key hyper-parameter -- embedding dimension -- to further investigate the MIRec performance on recommendation.

\subsubsection{TopK Performance} 
The top-K recommendation performance for all items, head and tail items are shown in Table \ref{table:perfmovie} for MovieLens1M and Table \ref{table:perfbx} for Bookcrossing. The user and item embedding dimensions for all methods are set to be the same for fair comparison. Overall, we see that MIRec generally brings the improvement in HR@K and NDCG@K for both tail and head items, and the overall performance. This is very encouraging since it's very hard to improve both tail and head items at the same time. Concretely, we have the following \textbf{key observations}:


\begin{itemize}

\item First, compared with the state-of-the-art two-tower model, the MIRec generally achieves great improvements for both tail items and head items. Since both methods are based on the two-tower architecture, the results verify the importance of considering the long-tail item distribution in recommendation prediction. The improvements of both head and tail items also confirm that the dual transfer learning framework indeed benefits the learning of items in the long-tail distribution for enhanced recommendation.



\item Second, among different strategies for long-tail distribution problem (e.g. re-sampling, re-weighting, and transfer learning), MIRec achieves the best performance. The results demonstrate the superiority of considering transfer learning among items in the long-tail item distribution. Concretely, as shown in Table \ref{table:perfmovie} and Table \ref{table:perfbx}, the re-sampling strategies (i.e. over-sampling and under-sampling) show poor performance in recommendation. The results are consistent with the previous findings \cite{yi2019sampling,wang2017learning} that re-sampling could heavily influence the model performance and cause sub-optimal performance due to the changes of the original data distribution. It also illustrates the importance of relatively keeping the original data distribution as it was in the training process. For the refining loss function strategies (e.g. logQ), they could achieve good performance for tail items, but perform quite neutral or poorly for head items. It is reasonable since logQ can give more weights to the tail items in the loss function. The results demonstrate the clear advantage of our dual transfer learning framework. In Table 5, for baselines that have better performance on tail items (e.g. LogQ and Head2Tail), their performance significantly decreases for head and overall performance. A possible reason is their learned feature representations are sub-optimal, causing an unhealthy trade-off between tail items and head items. Different from them, MIRec is able to achieve significant gains for tail items while maintaining or sometimes improving the head/overall item performance. This further shows the high-quality of learned representations by MIRec. 



\item Third, compared with the commonly used curriculum learning strategies (Head2Tail, Tail2Head) and state-of-the-art meta-learning (MeLu), MIRec obtains better performance. It shows by integrating curriculum learning with meta-learning, the dual transfer framework can better transfer knowledge in the long-tail item distribution. Note here the MeLu does not perform good, one possible reason is that it does not well fitted for the implicit feedback based recommendation. MeLu also requires a large amount memory when high dimensional item features are considered.


\end{itemize}
\begin{figure*}[t!]
\centering  \tiny
\vspace{-5pt}
 \includegraphics[width=6.3in]{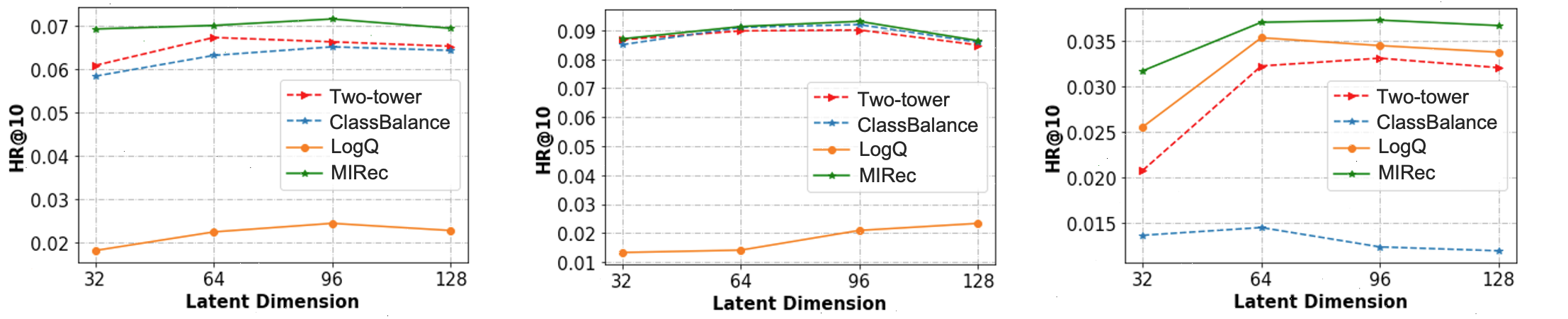}  
 \vspace{-10pt}
 \caption{Recommendation performance (overall, head and tail) for different latent dimensions on MovieLens1M dataset.}
  \label{fig:dim} 
 \vspace{-5pt}
\end{figure*}

\subsubsection{Influence of Embedding Dimension} We also analyze the key hyper-parameter – the embedding dimension – on MIRec to see its effect on MIRec's performance. Results for the MovieLens1M dataset is shown in Figure \ref{fig:dim}, with comparison of other representative methods. As shown in Figure \ref{fig:dim}, MIRec consistently outperforms the other methods when different embedding dimensions are considered. It further confirms the importance of considering long-tail item distribution and utilizing the transfer learning to learn the item relations in the long-tail distribution for improved recommendation. It is also worth noting that performance of tail items tends to degrade with higher embedding dimension. Our hypothesis is that tail items are easier to get overfitted due to having less data for training.

\subsection{Dual Transfer Learning (RQ2)} 
Given the great performance of MIRec, we further investigate how does each component in MIRec contribute to the improved performance on long-tail item recommendation. Specifically, we conduct an ablation study experimenting different versions of MIRec that only includes a specific component:

\squishlist

\item\textit{$MIRec_M$}: This only considers the meta-learning part in the MIRec. Similar as \cite{wang2017learning}, $\Omega^*$ and $\Omega(k)$ are constructed only with head items $i \in I_h(k)$. Concretely, $\Omega^*$ is built by all the user feedback of item $i \in I_h(k)$. We randomly sampling k user feedback of each $i \in$ $I_h(k)$ to construct $\Omega(k)$.


\item\textit{$MIRec_C$}: This only considers the curriculum learning part in MIRec. Specifically, the two-tower model is first trained based on $\Omega^*$, then fine-tuned on $\Omega(k)$.


\item\textit{$MIRec-LogQ$}: We use the $s(x_u, y_i)$ to calculate the user preference towards items instead of the logQ correction based $s^c(x_u, y_i)$ in Equation \eqref{eq:logq};


\item\textit{$Two-tower_{2}$}: MIRec uses both the many-shot model and meta-mapped model to predict the user preference (as shown in Equation \eqref{eq:pred}). Therefore, we use two two-tower models that have similar number of model parameters as MIRec for further comparison. That is, the prediction is based on the equation $s'(\mathbf{x}_u, \mathbf{y}_i) = \lambda_p g(\mathbf{x}_u,  \mathbf{y}_i; \theta^*) + (1 - \lambda_p)g(\mathbf{x}_u,  \mathbf{y}_i; \theta')$ where $\theta'$ is calculated by another two-tower model with different initialization. 
\squishend

The results of the ablation study is shown in Table \ref{table:abl}, that MIRec outperforms the other variations. This is to be expected since we expect all different components contribute to the improved performance.

Comparing $MIRec_C$ and $MIRec_M$, which only use one type of transfer learning, we see that $MIRec_M$ that uses the vanilla meta-learning method performs poorly. This indicates that directly applying the meta-mapping, which is learned on the head items only, could limit the learning of various tail item's representation. It also highlights the importance of considering curriculum transfer among items to make the learned meta-mapping suitable for tail items.

$MIRec_C$ achieves good performance for tail items, demonstrating the proposed curriculum (from $\Omega^*$ to $\Omega(k)$) indeed transfers knowledge from head to tail items to improve the performance of tail items. We also observe that $MIRec_C$ does not perform well on head items, which is expected. One hypothesis is that since we down-sample head items in the second phase, the model might have forgotten some of the patterns learned on head items.

Furthermore, comparing $MIRec-LogQ$ with MIRec, MIRec achieves a better performance, that verifies that logQ correction can further help the knowledge transfer in the long-tail item  distribution by adding the item frequency based correction. 

Lastly, for $Two-tower_{2}$, though it contains similar number of learned parameters as MIRec, $Two-tower_{2}$ performs poorly for tail items, compared with MIRec. One likely reason is that the combined two two-towel models in $Two-tower_{2}$ easily causes over-fitting for tail items. It further shows the importance of transfer knowledge in long-tail item distribution to improve the recommendation for tail items.

\begin{table*}[t!]
\vspace{-5pt}
\caption{Ablation study of MIRec on MovieLens1M.}
\vspace{-10pt}
\label{table:abl}
\begin{tabular}{c|cc|cc|cc}
\toprule
\multirow{2}{*}{Measure\%}           & \multicolumn{2}{c|}{Overall} & \multicolumn{2}{c|}{Head} & \multicolumn{2}{c}{Tail} \\ \cline{2-7}
          & HR@10        & NDCG@10       & HR@10      & NDCG@10      & HR@10      & NDCG@10     \\\hline
Default & 6.74         &  3.26             &   8.98         &  4.33            &   3.22         &    1.59         \\ \hline
$MIRec$     &  \textbf{7.02}            &     \textbf{3.36}          &    \textbf{9.13}       &     \textbf{4.36}       &    \textbf{3.70}        &    \textbf{1.81}       \\ 
$MIRec_M$  &    6.48          &      3.13         &   8.75       &    4.17          &   2.88        &    1.43       \\ 
$MIRec_C$  &  6.83            &     3.30          &     8.77     &     4.21         &    3.78       &    1.87       \\ 
$MIRec-LogQ$  &  6.98            &   3.36           &   9.75       &     4.71         &   2.62        &  1.26         \\ 
$Two-tower_{2}$  &  6.90            &   3.22            &    9.46      &    4.37          &   2.89        & 1.41           \\ \bottomrule
\end{tabular}
\end{table*}

\subsection{Curriculum Learning Process (RQ3)} 
The proposed curriculum learning -- Long-Tail distribution-aware CURriculum (LTCur) plays a pivotal role in transferring knowledge among head and tail items in MIRec. Thus, in this section, we further study the performance of curriculum learning strategies on both head and tail items in different curriculum stages. Specifically, we report the loss values on head \& tail slices, and overall performance over time.



\begin{figure*}[t!]
\centering  \tiny
\vspace{-5pt}
 \includegraphics[width=4.2in]{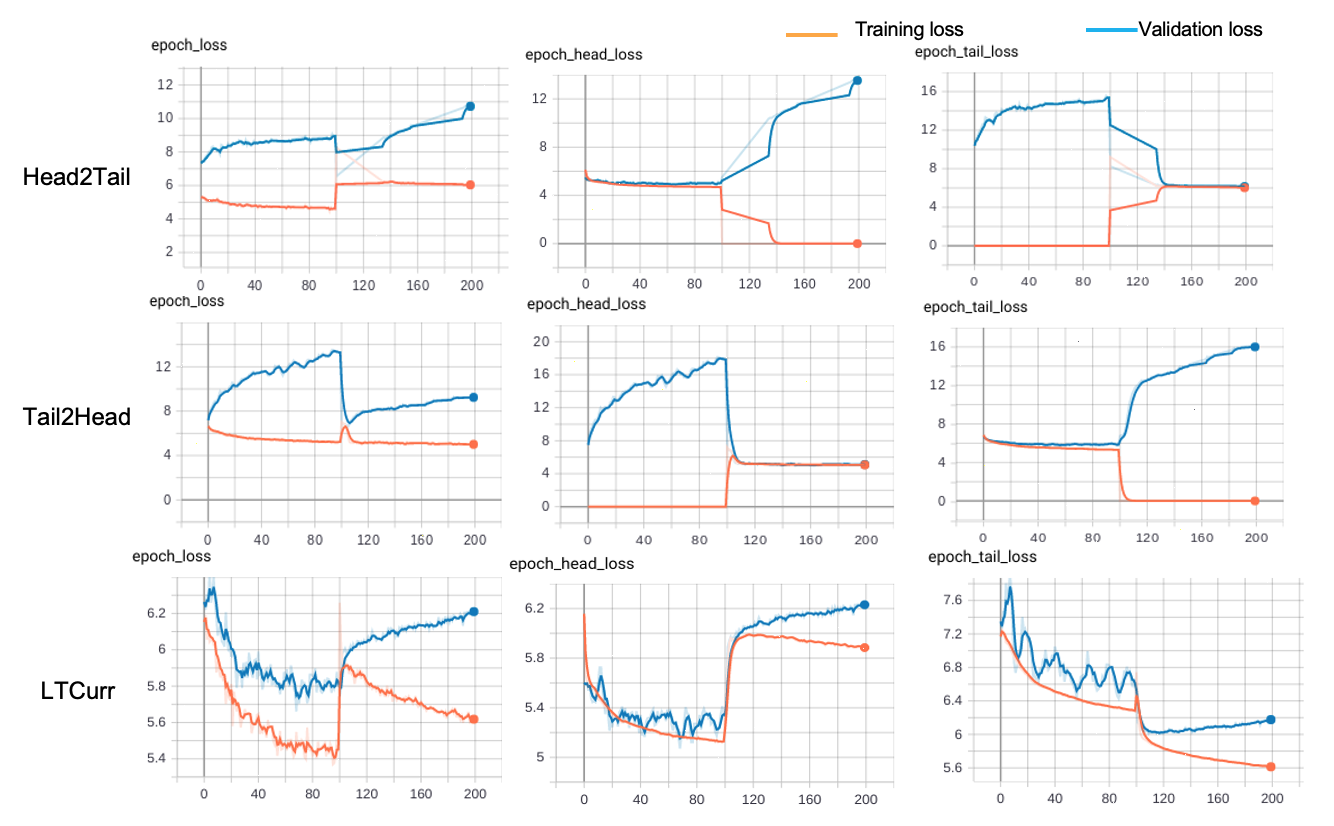}  
 \vspace{-10pt}
 \caption{The loss changes of different curriculums (by rows) for overall items, head items and tail items (by columns) on MovieLens1M. x-axis is the index of epoch. Blue/red line shows the loss of validation/training dataset. }
  \label{fig:loss} 
 \vspace{-5pt}
\end{figure*}

In Figure \ref{fig:loss}, each row represents a curriculum strategy. x-axis represents the epoch index, and  y-axis reports loss values. Each curriculum contains two curriculum stages. Each line is \textit{smoothed}. The training loss values (red line) is 0 if the head/tail items are not trained. For example, in Head2Tail, since the tail items in the training dataset are not trained, the training loss of tail items (row 1 column 3 in the red line) is 0 in the first curriculum stage, similar for the head items in the second stage (row 1 column 2 in the red line). 


From the Figure \ref{fig:loss}, we observe that: (1) Compared to the tail item loss in different curriculums (column 3), our proposed curriculum can bring a two-stage decent for both the training and validation loss, as shown in row 3 column 3 in Figure \ref{fig:loss}. It demonstrates the LTCur curriculum enhances the learning of tail items in both two curriculum stages. (2) When the model is trained based on only head/tail items, the validation performance for the other part of items decreases. The different changes of head and tail loss indicate the large variations between head and tail items, which further confirms the importance of transferring knowledge between head and tail items for enhanced recommendation. (3) It is easily to get validation loss increases if the model is trained purely based on head/tail items. As shown in first column of the first two rows: when the training loss decreases (red line), the validation loss would increase (blue line). It highlights the importance of considering both head and tail items in different curriculum stages. 






\begin{figure}[t!]
\centering  \tiny
\vspace{-10pt}
 \includegraphics[width=3.2in]{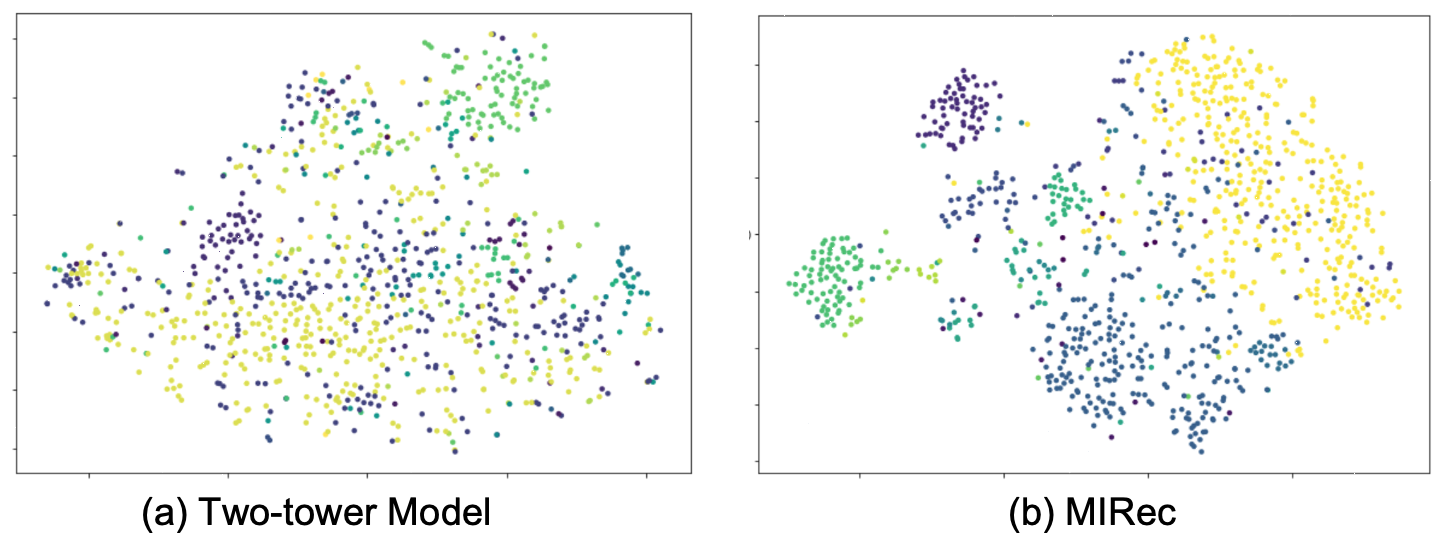}  
\vspace{-10pt}
 \caption{The 2-D visualization (with t-SNE \cite{van2014accelerating}) of tail items on MovieLens1M. The color represents the movie genres. }
  \label{fig:visual} 
\end{figure}


\subsection{Embedding Visualization and Case Study (RQ4)} 
In this section, we look into the learned representations from MIRec by visualizing the embeddings, and examine the performance of MIRec qualitatively.



First, we visualize the learned tail item embeddings from the baseline (two-tower model) and MIRec using t-SNE (Figure \ref{fig:visual}) and highlight movie genres by different colors. We observe that the movie clusters from MIRec, compared with the baseline model, is more coherent w.r.t. the movie genres information. This suggests that the improved performance from MIRec can be attributed to better capturing the semantic information of tail items.

\begin{figure}[t!]
\centering  \tiny
 \includegraphics[width=3.1in]{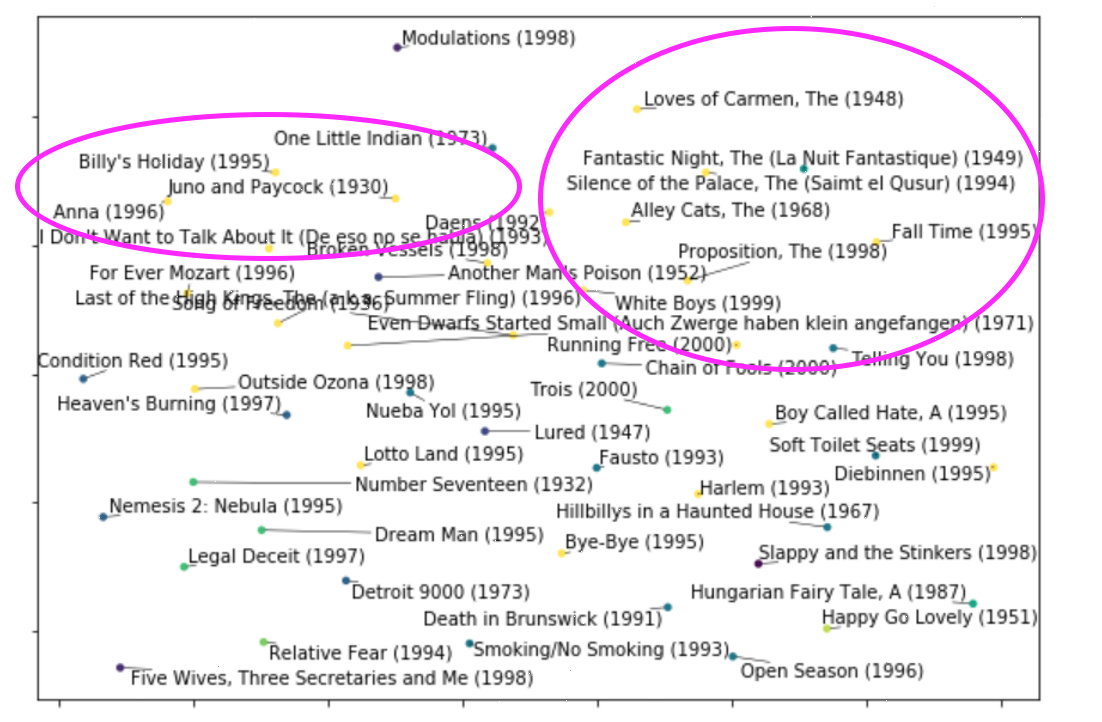}  
\vspace{-10pt}
 \caption{Case study: movies are clustered together by MIRec. The two purple circles highlight movies with similar themes are close to each other in the embedding space learnt by MIRec.}
  \label{fig:case} 
\end{figure}

Next, we further look into what movies that MIRec clusters together and visualize the clusters by movie titles in Figure \ref{fig:case}. Here we observe similar movies are clustered together. For example, `Billy's Holiday' and `Juno and Paycock', which are both about people's life, are clustered together. The visualization shows that although the tail items contain less collaborative information, MIRec can still well cluster those tail items based on their content information. This further validates our previous findings on MIRec.



\section{Conclusion}
In this work, we tackle the challenge of recommendations in the context of long-tail item  distribution. We propose a dual transfer learning framework -- MIRec -- that integrates both (i) meta-learning to transfer knowledge among models, and (ii) curriculum learning to transfer knowledge among items. The proposed curriculum learning inside the meta-learning ensures the smooth knowledge transfer from head items to the large number of tail items in recommendation. Extensive experiments on the two benchmark datasets show that MIRec consistently outperforms the state-of-the-art algorithms. For future work, we are interested in (i) improving the curriculum learning inside the meta-learning approach, and (ii) exploring other types of content information, such as incorporating item knowledge graph with the long-tail item  distribution.

\begin{acks}
The authors would like to thank Maciej Kula, Jiaxi Tang, Wang-Cheng Kang for their help in this work.
\end{acks}

\bibliographystyle{ACM-Reference-Format}
\balance
\bibliography{ref}


\end{document}